# On the Origins and Relevance of the Equal Transit Time Fallacy to Explain Lift

Graham Wild
School of Engineering and Information Technology, UNSW ADFA, Canberra, Australia

G.Wild@ADFA.edu.au

**Abstract**

Recently, aerodynamics syllabi have changed in high schools, pilot ground training, and even undergraduate physics. In contrast, there has been no change in the basic theory taught to aeronautical or aerospace engineers. What has changed is technology, both experimentally and computationally. The internet and social media have also empowered citizen science such that the deficiencies in the legacy physics education around flight and lift are well known. The long-standing equal transit time (ETT) theory to explain lift has been proven false. If incorrect, why was it ever taught? Through a historical analysis of relevant fluid and aerodynamics literature, this study attempts to explain why ETT theory is part of our collectively lower-level cognitive understanding of lift and flight. It was found that in 1744 D'Alembert himself assumed this to be a feature of moving fluids, and while this initial intuition (ETT 1.0) was incorrect, the property of ETT (ETT 2.0) was derived in 1752 when applying Newton's laws of motion to fluids. This incorrect result was independently confirmed in 1757 by Euler! The conclusion is that an over simplified treatment of fluids predicts ETT, along with no lift and drag. This then leads to the open question, can ETT be taught at an appropriately low level as an explanation for lift?

## Introduction

Over the past several years, ubiquitous computing has seen two key developments relevant to a discussion on *misconceptions* in the explanation of lift. The first is the growth of computational fluid dynamics (CFD). The second is online science content creators using video-based social media. These two elements have conspired together to ensure the *myth* of equal transit time (ETT) associated with lift is well and truly *busted*. Recently, science communication channels on YouTube have attempted to address the issue of lift (see for example Veritasium and Sixty Symbols). YouTube also has footage utilising pulsed smoke flow visualization *proving* the air over the upper surface is much faster (see the Cambridge University channel) (Babinsky, 2003). While the author asked his own high school teacher in 1995 the dreaded question *why*, little effort appears to have been dedicated to the higher-level question, pedagogically speaking. The question to be answered is, "why did anyone ever teach ETT to begin with, if it is wrong?" While ETT is incorrect, people at some point in time did not think so, or it would not be offered as part of an explanation of lift.

It is worth noting that there are proponents of modern origins to the ETT fallacy. Two great names are implicated, Lilienthal (1889) and Prandtl (1921). The evidence for these indictments are a statement that air flowing around a body is "mainly confined to opening out in front of the advancing body and joining up again in its rear" (Lilienthal, 1889), and an image that could imply ETT (Prandtl, 1921). The evidence of Weltner and Ingelman-Sundberg (2003) is circumstantial at best. Lilienthal does not suggest anything more than the flow will be smooth, continuous, and attached (not stalled). Similarly, Prandtl's diagram shows the upper flow moving faster on closer inspection. The hypotheses of modern origins ignore additional information on the nature of ETT in fluid mechanics and aerodynamics. As such, further investigation is necessary to understand the true origins of ETT.

**Equal Transit Time**

First, for those unfamiliar, what is ETT? Consider a simple wing cross section, an aerofoil, the most basic geometry is a circular segment from a chord, shown in Fig. 1, analogous to that shown by Zetie (2003). Also shown in Fig. 1 are streamlines, as well as particles traveling along them shown at seven different time increments (t = 0 to 6). The key feature is how each particle is displaced an equal distance along the flow at the same time. The result of the longer path above combined with the equal time, means a faster flow velocity. Bernoulli's principle is then invoked to convert the higher velocity to a lower pressure (Bernoulli, 1738), giving a pressure difference and hence lift.

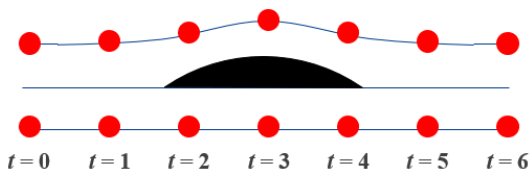

Fig. 1: The incorrect flow of a fluid around a segment from a chord illustrating equal transit above and below.

A potential source of the idea of ETT may come from incorrect assumptions about the flow. Specifically, if you consider a streamline a sufficiently large distance above and below, pressure *information* from the passage of the aerofoil cannot reach these points in the fluid. One may further reason that this information (energy) must also diminish, resulting in less curvature further away. If you then only consider this far field, all particles above and below transit in equal time. This reasoning appears to the that used by D'Alembert (1744), who stated that uniform vertical slices of fluids would be parallel; this form of intuitive ETT will be denoted as ETT 1.0. This appears analogous to optical wavefronts from Huygen's Principle. As mentioned, we can experimentally prove D'Alembert's assumption is incorrect; hence, the common thinking associated with this is that CFD and high-speed cameras (for flow visualization) being modern developments could only recently show ETT to be a fallacy.

**Potential Flow**

An infamous point in the development of fluid mechanics is D'Alembert's paradox. D'Alembert's study of the wind and fluids famously concluded with a prediction of zero drag (and lift) experienced by a body in a fluid (Spalart, 2015), and hence the use of the term "paradox". The initial abstract mathematic fluid utilised by early theoretical fluid dynamicists is called potential flow (Anderson, 2011). This was natural, given gravitational and electromagnetic fields are defined by equivalent mathematical potentials (O'Neil, 2017). As they are derived from solutions of Laplace's equation, potential flows are incompressible, inviscid, and irrotational. That is, they have constant density, possess no friction (viscosity), and their vorticity and circulation are zero. Two useful features of potential flows are superposition and conformal mapping. These features allow flow field solutions to be combined, creating more complex flows, and for simple geometric features to be transformed to more complex geometries. For a visual introduction, Dr Tom Crawford from the University of Oxford has a YouTube video in collaboration with Grant Sanderson, called "Potential Flow and Method of Images with 3Blue1Brown" (Crawford, 2021).

Consider the combination of "free stream" uniform flow and a doublet (Fig. 2). This linear combination gives an object immersed in the potential, with the *fluid* flowing around. The observed boundary at the edge of the object is called a Rankine Oval (Katz and Plotkin, 2001). The simple analogy here, is that of a compass (a magnetic doublet, or a dipole) in the earth's magnetic field at the surface (the uniform field). As is commonly known, the compass needle aligns with the external magnetic field.

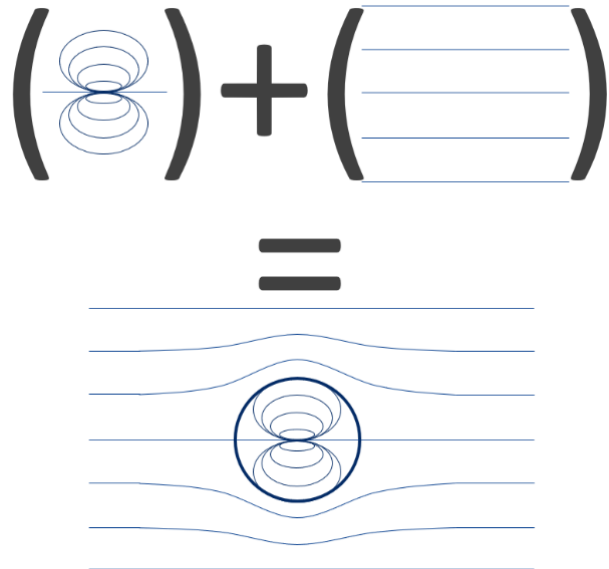

Fig. 2: Potential flow superposition of a uniform flow and a doublet, resulting in a flow "around" an "object".

Since the superposition of a uniform flow, a source, and a sink is a valid solution, it can be conformally mapped to a new shape. This will produce the potential flow for the new geometry created through the conformal mapping. Due to their mathematical simplicity (Matthews, 2012), the conformal mapping typically utilised is the Joukowski transformation, producing Joukowski aerofoils (Joukowski, 1910). However, since lift is also generated by a flat plate (equivalent to that of a thin symmetric aerofoil at the same angle of attack), a conformal mapping to a flat plate is of greater benefit, conceptually (Fig. 3). It should be noted that this is also mathematically possible using a Tsien transform (Tsien, 1943). Rather than utilising mathematics, with care, logical reasoning can be used to understand the resultant flow. Mapping the potential flow solution for a cylinder to a plate flat in y (no height), the flow field returns to just the free stream case, noting stagnation points at the leading edge (LE) and trailing edge (TE). Mapping to a plate flat in x (no width), the solution is also intuitive; as with the cylinder, the flow is forced around the plate with the stagnation points at the centres of the front and rear surfaces.

for the given transformation must be rotationally symmetric. Therefore, any corresponding changes in static and dynamic pressure over the upper surface are also required to be rotationally symmetric relative to those over the bottom surface. This is a consequence of the conformal mapping and irrotational nature of the flow, which requires that the circulation is zero. The aerodynamic result of potential flow is a pitching moment with no lift (or drag). Importantly, the consequence of pure potential flow is equal transit! This consequential ETT will be denoted as ETT 2.0, in contrast to D'Alembert's previously discussed intuition (ETT 1.0).

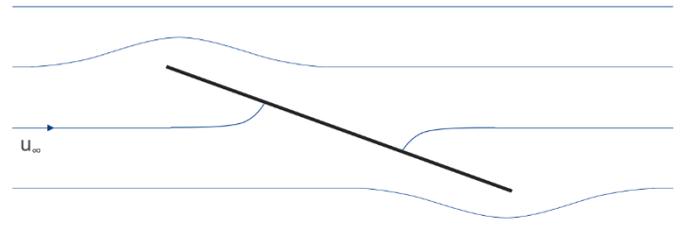

Fig. 4: Conformal mapping of potential flow to a plate inclined at 30 degrees.

**Circulation**

Clearly, since the very nature of D'Alembert's paradox is that potential flow produces no lift or drag, this solution which *predicts* ETT is of little aerodynamic significance, and arguably no aerodynamic use. To productively use potential flow along with a Joukowski transformation the paradox needs to be resolved; that is, the flow needs to be *fixed*.

In undergraduate aerodynamics classes, students are taught from the beginning to add circulation to potential flow to facilitate lift calculations (Anderson, 2015). This *fix* is the Kutta-Joukowski theorem and includes the important Kutta condition, which is required to force the rear stagnation point to the TE. The *fix* is valid, given that the stagnation point is located at the TE for real viscous fluids. The result of this for aerospace/aeronautical engineers (student or otherwise), is skipping over the 150-year gap in aerodynamics development (1750 to 1900), the *aerodynamic dark ages*. Engineers go straight to a working understanding of lift, looking at solutions to Laplace's equation, for potential flow, and adding circulation. As such, these engineers who are most able to comprehend the existence the ETT fallacy, are instead perplexed as to why "no one else gets it" (*it* being lift). This also explains why *circulation* as a theory for explaining lift is growing in popularity. Although, it should be noted in the realm of cause and effect, circulation is an effect, not a cause (McLean, 2012, Anderson, 2015), unlike the somewhat analogous Magnus effect (lift on a spinning ball). That is, circulation gives the correct value for lift; however, it does not address why the circulation, and a specific amount, exists (other than for theory to match experiments). In other words, the consequence of circulation around an aerofoil is that the

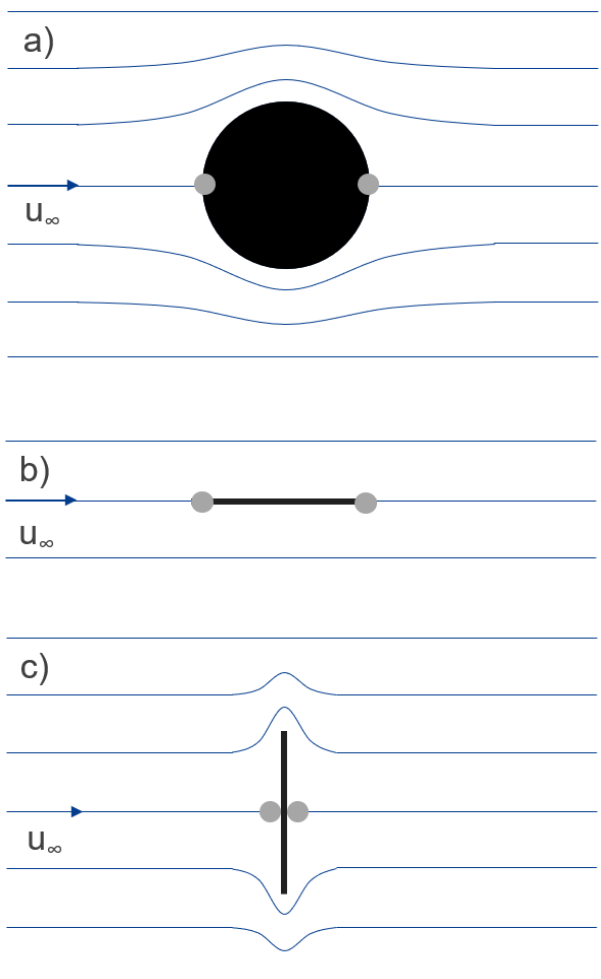

Fig. 3: Conformal mapping of the potential flow from around a sphere (a) to a plate flattened vertically (b) and horizontally (c).

If the potential flow is mapped to a flat plate at any other angle, the conformality (angle preserving nature) of the transformation requires the stagnation points to be mapped perpendicular to the surfaces. At a positive angle of attack, the rear stagnation point on a plate in a potential flow is mapped to the upper surface, as shown in Fig. 4. The resultant solution flow must be faster over the upper surface relative to the lower surface, such that the relative difference in pressure can be logically derived. Similarly, the fact that fluid is displaced downwards (the flow is no longer symmetric) means a conservation of momentum argument can be used. Both these two previous points about circulation (difference in velocity and hence pressure, and the displacement of fluid downwards) explain *how* lift is generated; however, they do not address *why*.

**Navier-Stokes**

In contrast to circulation, the Navier-Stokes equations (laws of motion) for fluids explain not only *how* but *why* an aerofoil produces lift. The reason potential flow with circulation is taught and utilised is because it is an analytical solution, giving good approximation for the flow velocity and pressure, as well as the resultant lift force. In contrast, solving Navier-Stokes equations typically requires the use of computers and numerical techniques, CFD.

The exclusive use of circulation also hides another important feature and likely explains why even aero engineers struggle to understand the persistence of ETT. This fact is the hidden paradox in Navier-Stokes (Gonzalez and Taha, 2021). The paradox comes from the fact that an inviscid flow has zero viscosity. However, viscosity is inversely proportional to Reynold's number for a fluid. So as viscosity tends to zero, Reynold's number tends to infinity, and potential flow should result (specifically, the inviscid Euler equations). However, when visualised, it is not flows with very high Reynold's numbers that appear similar to potential flow; in fact, as seen in Fig. 5 it is flows with very low Reynold's numbers (Feynman et al., 2010 [1964]), that is, very high viscosities.

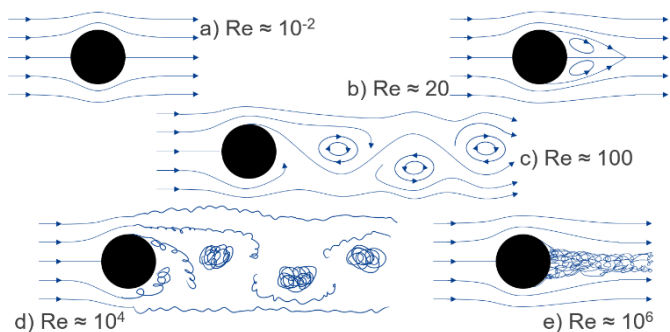

Fig. 5: Flow past a cylinder for various Reynolds numbers. Adapted from Feynman et al. (2010 [1964]).

Richard Feynman famously differentiated potential flow and real fluids as dry and wet water, respectively (Feynman et al., 2010 [1964]). The apparent paradox in Navier-Stokes is resolved by Feynman et al. (2010 [1964]), arising from the product of viscosity and the Laplacian (the divergence of the gradient, or somewhat analogous to a second derivative) of the flow velocity. These conspire together such that a turbulent flow with a low viscosity is dominated by significant changes in the second order derivative. That is, if viscosity is vanishingly small, the analogous jerk quantity will tend to infinity (noting that jerk is the second derivative of velocity). The result of this and the complexity of Navier-Stoke means that even after a correct and complete theory of fluids existed, which was capable of quantifying drag and lift, there was renewed interest in inviscid theories; specifically, inviscid approximations with corrections to enable potential flow to be used as extensively as possible, given the simplicity in the mathematics. Prandtl (1904) facilitated this with the no-slip condition and the concept of the boundary layer, wherein almost all viscous effects occur. This in turn leads to *circulation* as an approximate solution to facilitate calculations, albeit with the previously mentioned limitations vis-a-vis *how* and *why*.

**Pedagogical Implication**

At this point, consider a parallel educational question; is Newton's Law of Universal Gravitation incorrect? While a fundamental aspect of physics being taught in secondary and tertiary classes (Cunha and Tort, 2017), it is hoped that most are aware of the limitation of Newtonian gravity (Poisson and Will, 2014). However, a quick search of the news media makes it fairly clear that many in the general public are unaware, see for example Fish (2019). While it is hard to think that this cornerstone of physics will ever disappear from high school textbooks, could it be relegated to primary schools, if not skipped all together? The power of computers and the changing nature of education and syllabi means concepts once considered "hard" can become trivial (especially with the aid of technology). General relativity can be part of a secondary curriculum (Stannard, 2018), or even a primary one (Ruggiero et al., 2021); where does that leave Newton?

Is the story if ETT analogous to other aspects of our knowledge and understanding, like gravity? That is, is ETT like Newtonian Gravity, with Navier-Stokes being the General Relativity of aerodynamics? While never correct, the fact that ETT has persisted until recently clearly shows that the intuitions and results of D'Alembert (and Euler) are an important part of our collective understanding of aerodynamics. Similarly, better theories and understandings have effectively *deleted* the concept from those with a greater understanding of aerodynamics. Furthermore, CFD for flow visualization around aerofoils means Navier-Stokes calculations are no longer too complex.

**Conclusion**

ETT is the result of D'Alembert's (1752) solution, along with an equivalent analysis by Euler (1757); that is, it is a result of pure potential flow. While of little aerodynamic significance, does ETT belong in primary or secondary education? This question remains, with no definitive objective answer, unlike the question to the origins of the ETT theory. The focus of aerodynamicists and aerospace engineers is that a theory of lift needs to have lift as a by-product, and ETT does not. Then again, maybe a theory of gravity needs to account for general relativistic effects, and Newton does not. The result of the authors 25 year-long quest to understand ETT is an understanding of the historical constructivist nature of physics knowledge. This has been set out herein, answering the question, "why is ETT theory part of our collectively lower-level cognitive understanding of lift generation?"

At this point, a key recommendation can be made. While learning about D'Alembert's paradox, aeronautical and aerospace engineering students should be made aware that ETT as a *high school* explanation of lift (old or current), is a direct result of pure potential flow, and hence why it was assumed true. Similarly, one would hope that Feynman's approach to teaching fluids may influence undergraduate physics courses, such that a better understanding of fluids and aerodynamics is possible.

With a suitable hypothesis presented on the origins of the ETT theory, it is hoped that this knowledge can be utilized by educationalists when deciding what a science syllabus should contain. Specifically, if the concept of ETT may be utilised at all in explanations of lift at suitably low levels. For the sake of completeness, a correct detailed explanation of lift is provided in the works of retired Boeing aerodynamicist Doug McLean (McLean, 2018a, McLean, 2018b, McLean, 2012).

**References**

ANDERSON, J. D. 2011. *Fundamentals of Aerodynamics,* New York, NY, McGraw-Hill Education.
ANDERSON, J. D. 2015. *Introduction to Flight,* Boston, MA, McGraw-Hill Education.
BABINSKY, H. 2003. How do wings work? *Physics Education,* 38**,** 497-503.
BERNOULLI, D. 1738. *Hydrodynamica [Hydrodynamics],* Germany, Johannis Reinholdi Dulseckeri (Publisher).
CRAWFORD, T. 2021. Potential Flow and Method of Images with 3Blue1Brown. YouTube.
CUNHA, R. F. F. & TORT, A. C. 2017. Plausibility arguments and universal gravitation. *Physics Education,* 52**,** 035001.
D'ALEMBERT, J. 1744. *Traité de l'équilibre et du mouvement des fluides [Treatise on balance and movement of fluids],* Paris, France, Michel-Antoine David (Publisher).
D'ALEMBERT, J. 1752. *Essai d'une nouvelle théorie de la résistance des fluides [Testing of a new theory of fluid resistance],* Paris, France, Michel-Antoine David (Publisher).
EULER, L. 1757. Principes généraux du mouvement des fluides [The General Principles of the Movement of Fluids]. *Mémoires de l'académie des sciences de Berlin [Journal of the Berlin Academy of Science],* 11**,** 274-315.
FEYNMAN, R., LEIGHTON, R. & SANDS, M. 2010 [1964]. The flow of wet water. *Feynman Lectures on Physics,* 2**,** 41.1-41.12.
FISH, T. 2019. Newton was wrong: Scientists dismiss Newton's theory of gravity and warn Einstein is next. *Express (Online)*, Aug 13.
GONZALEZ, C. & TAHA, H. E. 2021. Are Superfluids Lifting? A Novel Variational Theory of Lift. arXiv.
JOUKOWSKI, N. E. 1910. Über die Konturen der Tragflächen der Drachenflieger [About the contours of the wings of the hang-glider]. *Zeitschrift für Flugtechnik und Motorluftschiffahrt [Magazine for aviation technology and motorized airship travel],* 1**,** 281-284.
KATZ, J. & PLOTKIN, A. 2001. *Low-speed aerodynamics,* Cambridge, UK, Cambridge University Press.
LILIENTHAL, O. 1889. Birdflight As the Basis of Aviation, translated by AW Isenthal. *Hummelstown, Penn.: Markowski International.*
MATTHEWS, M. T. 2012. Complex mapping of aerofoils – a different perspective. *International Journal of Mathematical Education in Science and Technology,* 43**,** 43-65.
MCLEAN, D. 2012. *Understanding Aerodynamics: Arguing from the Real Physics,* West Sussex, UK, Wiley.
MCLEAN, D. 2018a. Aerodynamic Lift, Part 1: The Science. *The Physics Teacher,* 56**,** 516-520.
MCLEAN, D. 2018b. Aerodynamic Lift, Part 2: A Comprehensive Physical Explanation. *The Physics Teacher,* 56**,** 521-524.
O'NEIL, P. V. 2017. *Advanced Engineering Mathematics,* Boston, MA, Cengage Learning.
POISSON, E. & WILL, C. M. 2014. *Gravity: Newtonian, Post-Newtonian, Relativistic,* Cambridge, UK, Cambridge University Press.
PRANDTL, L. Über Flussigkeitsbewegung bei sehr kleiner Reibung [Via fluid movement with very little friction]. International Congress of Mathematicians, Aug 8-13 1904 Heidelberg, Germany. 484-491.
PRANDTL, L. 1921. Applications of modern hydrodynamics to aeronautics. *US Government Printing Office.*
RUGGIERO, M. L., MATTIELLO, S. & LEONE, M. 2021. Physics for the masses: teaching Einsteinian gravity in primary school. *Physics Education,* 56**,** 065011.
SPALART, P. R. 2015. Extensions of d'Alembert's paradox for elongated bodies. *Proceedings of the Royal Society A: Mathematical, Physical and Engineering Sciences,* 471**,** 20150106.
STANNARD, W. B. 2018. Why do things fall? How to explain why gravity is not a force. *Physics Education,* 53**,** 025007.
TSIEN, H.-S. 1943. Symmetrical Joukowsky airfoils in shear flow. *Quarterly of Applied Mathematics,* 1**,** 130-148.
WELTNER, K. & INGELMAN-SUNDBERG, M. 2003. Physics of Flight–reviewed. *submitted to Eurpean Journal of Physics.*
ZETIE, K. 2003. Teaching about wing lift. *Physics Education,* 38**,** 486-487.